# Digital contact tracing/notification for SARS-CoV-2: navigating six points of failure

Joanna Masel[1,*], James Petrie[2,3], Jason Bay[4], Wolfgang Ebbers[5], Aalekh Sharan[6], Scott Leibrand[7], Andreas Gebhard[8], Samuel Zimmerman[9]

[1] Dpt. Ecology & Evolutionary Biology, University of Arizona, Tucson AZ, USA
[2] Dpt. Applied Math, University of Waterloo, Waterloo Ontario, Canada
[3] present address: Big Data Institute, University of Oxford, Oxford U.K.
[4] TraceTogether, Singapore
[5] Dpt. Public Administration and Sociology, Erasmus University Rotterdam, the Netherlands
[6] Aarogya Setu, India
[7] CoEpi, Seattle WA, USA
[8] TCN Coalition and Forward Momentum, LLC, New York NY, USA
[9] PathCheck, Cambridge MA, USA
[*] corresponding author

## Abstract

Digital contact tracing/notification was initially hailed as a promising strategy to combat SARS-CoV-2, but in most jurisdictions it did not live up to its promise. To avert a given transmission event, both parties must have adopted the tech, it must detect the contact, the primary case must be promptly diagnosed, notifications must be triggered, and the secondary case must change their behavior to avoid the focal tertiary transmission event. If we approximate these as independent events, achieving a 26% reduction in R(t) would require 80% success rates at each of these six points of failure. Here we review the six failure rates experienced by a variety of digital contact tracing/notification schemes, including Singapore's TraceTogether, India's Aarogya Setu, and leading implementations of the Google Apple Exposure Notification system. This leads to a number of recommendations, e.g. that tracing/notification apps be multi-functional and integrated with testing, manual contact tracing, and the gathering of critical scientific data, and that the narrative be framed in terms of user autonomy rather than user privacy.

## Introduction

Contact tracing is a time-tested tool to fight an emerging outbreak of infectious diseases such as SARS-CoV-2. If x% of infections are identified, and y% of those in contact with a known case are traced in time and persuaded to stay home, then in a well-mixed population, the effective reproduction number R(t) will go down by a factor of x*y. This seems sufficient to reach the level R(t)<1 needed to quash an outbreak of many emerging pathogens (which have not yet evolved $R_0$>>1) before the outbreak develops too far, especially in combination with modest social distancing. However, SARS-CoV-2 proved to be a particular challenge for contact tracing. With short incubation periods and presymptomatic transmission making it difficult to trace contacts in time, digital contact tracing held tremendous promise, especially when leveraging smartphones that were already in consumers' hands. Here we reflect on why digital alternatives to traditional contact tracing instead had limited impact on SARS-CoV-2 transmission. We do so with a focus on Exposure Notification and related protocols, in part by drawing on first-hand experiences from our various involvements with this technology, including material that has not previously been public, as well as publicly available documents that are not indexed for literature searches. Our intent is to better inform those who might wish to prepare for and/or fight a new pandemic with a similar technological approach, allowing them to learn from what happened during the COVID-19 pandemic.

A brief note on terminology. Some digital protocols make it impossible to identify pairs of interacting individuals, even in cases where one transmitted disease to the other - in this case we refer to "notification" instead of "tracing". We refer to "proximity" vs. "presence" notification/tracing on the basis of whether exposure is assessed on the basis of proximity to an individual vs. presence at a shared-air venue. We reserve the term Exposure Notification (EN) for the specific protocol implemented by Apple and Google; this protocol is a form of proximity notification. EN is one of several protocols to assess proximity on the basis of signals sent and received between pairs of devices using low energy Bluetooth. Other protocols may use ultra-wideband or ultrasound for proximity detection, and/or detect presence from other information such as QR code scan histories, GPS coordinates, or logs of Wi-Fi access points. These technologies can also be used to warn of future infection risk, rather than trigger quarantine, testing, and isolation among those already infected [1] - this alternative purpose is out of scope here.

To be effective in stemming transmission, a notification must navigate six potential points of failure:
1. The primary case must have the tech in place at the time of transmission
2. The secondary case must have the tech in place at the time of transmission
3. The exposure that resulted in transmission must be judged to be high risk
4. The primary case must obtain a positive diagnosis in a timely manner
5. Notifications stemming from the primary case must be rapidly triggered following a positive diagnosis

6. After receiving a notification, the secondary case must change their behavior in a manner that prevents onward transmission to tertiary cases

If each of these steps were successful 80% of the time, and we approximate the six steps as independent events, then transmission (R(t)) would be reduced by $0.8^6$ = 26%. While not transformational on its own, this would be a significant contribution to quashing an outbreak, and non-independence will make this figure somewhat higher. But if each step were successful a still-respectable 40% of the time, again assuming independent events, R(t) is reduced by only 0.4%. While this can make a valuable contribution to flattening a curve [2], or to reducing the stringency of indiscriminate social distancing [3], it falls far short of containing a pandemic. Given this simple mathematical consideration, if the aim is containment such that life is relatively normal while waiting for a vaccine, then we clearly need to achieve low failure rates at each one of the six failure points.

We next discuss each failure point in turn. We give some history of how they were handled, together with speculation about how they might have been handled better.

## 1-2. App adoption

We jointly discuss the first two points of failure, because they are so similar; both involve tech adoption among the population destined to become infected. Following media coverage of the model of Ferretti et al. [4], there was more attention on app adoption than on other points of failure, with an odd obsession with 60% adoption in the general population as a magic number [5]. Note that the adoption rate that matters is not that of the general population, but of the population likely to become infected. For SARS-CoV-2, this meant that during stay-at-home mandates what mattered was adoption among essential workers; when staying at home was common but optional, young adults constituted a disproportionate share of cases. However, app adoption among the general population can serve as a rough approximation.

The most straightforward way to achieve low failure rates at points #1 and #2, while still maintaining user autonomy, is to make the broadcasting and reception of appropriate signals opt-out for all smartphones. Opt-in consent of the primary case would still be required at failure point #5, and that of the secondary case at #6. Apple and Google were the only entities with the power to make an opt-out implementation happen, with competitors like Huawei having little market share. Despite a secure EN design that ensures that no information about Bluetooth signals sent or received leaves the phone without subsequent user consent, they chose to make Bluetooth broadcasting and receiving opt-in. We note that Apple's "Find my…" service uses Bluetooth in a similar fashion, and is at the time of writing opt-out rather than opt-in [6].

An even lower failure rate could have been achieved by a protocol proposed by one of us [JP] to use the Wi-Fi logs already collected by iOS and Android operating systems instead of an additional, purpose-built Bluetooth layer. Had Google and Apple chosen to allow an app post-hoc access to these logs, or to build OS functionality around them, this would have completely eliminated failure point #1 for individuals who regularly carry smartphones. Individuals would still need to have the tech switched on in order to bypass failure point #2 to receive a notification. As with Bluetooth solutions, those who do not regularly carry smartphones (mostly young children and the elderly) would need to be provided with other devices that could record proximity or presence; in Singapore, fobs covered 20% of the population that would not otherwise have been able to participate.

In understanding why Apple and Google did not make the tech opt-out or Wi-Fi based, it is important to consider the incentive structure from Apple's and Google's point of view. Co-operation requires time from their employees, and has no apparent commercial upside (beyond the economic gains to the company should efforts succeed in containing a pandemic). From their perspective, the primary consideration is the effect of their actions on brand perception. The upside impact on their brand is gained primarily from doing "something" (e.g. working together), but there is no proportionate impact from doing something more rather than less effective. The potential for downside impact is substantial, and focused mostly on privacy concerns, exacerbated by any ceding of control over the technology and its marketing to public health. They are thus incentivized to do "something", to carefully manage perceptions of that something, and to avoid brand risk, but not to increase effectiveness against disease transmission. In future, if companies like Apple and Google are to be gatekeepers of such tech, governments should devote urgent attention to aligning their incentive structure.

Achieving high opt-in adoption was hard. Singapore's non-EN solution succeeded, with opt-in adoption above 90%, even before the app became required for entry to public spaces [7]. In contrast, the highest EN adoption rates are of Germany's Corona Warn App, used by ~34% of the population [8], and the National Health Service (NHS) COVID-19 app, with 17-25% of the population of England and Wales having activated Bluetooth exchanges activated in 2020 and 2021 [9]. Lower rates were reported elsewhere: 8% in Canada, 13% in New Zealand, 10-16% in the Netherlands, and 19% in Switzerland [10].

A positive caveat to this fairly bleak assessment of uptake is that if one individual has adopted the app, it is more likely that other members of their social network have too. In other words, failures at points 1 and 2 are correlated, making the probability of overall failure lower than it would be if each failure were an independent event [11]. Both for this reason, and as part of good marketing practices [12], it can make sense to look at adoption rates within smaller communities. Adoption of 46% among cases was achieved in a campus setting with an intensive marketing push [13], and over 33% of the population in an island setting [14].

Key to the relatively high adoption of the NHS COVID-19 app and Corona Warn app were their additional functionalities. When entering a venue, scanning a QR code with the NHS COVID-19 app was offered as an alternative to writing down contact details. While subsequent presence tracing on the basis of either form of information was conducted relatively rarely (see section below), the requirement to check in to a venue did prompt app installation [15], especially in the subset of the population most likely to become infected through contact with strangers. The NHS COVID-19 app was also useful for ordering tests, and both it and the Corona Warn App offered a rapid and secure system for receiving test results. Later, the integration of EU digital COVID certificates (for vaccination, recent negative test result, or recovery from infection) triggered another wave of app adoption. Near-universal adoption in Singapore was achieved only after the proximity tracing app became the only means for compulsory venue check-in [16]. Similarly, adoption rates in India skyrocketed after the government made the app mandatory in order for a smartphone owner to move freely in public areas. We believe that to achieve high adoption, future pandemic apps need to seamlessly integrate services that are useful and convenient to users.

In late 2020, Apple and Google launched "Exposure Notification Express" (ENX). In Androids, this is a simple app that public health authorities can auto-generate from a small set of choices. In iOS, it is part of the Settings. Apple claimed that placing the scheme in the Settings created less friction for opt-in than was present for an app, and hence that ENX should be preferred to a custom app. However, ENX never reached usage levels comparable to the more successful European EN apps. This is despite the fact that the phones of residents of adopting US States were pinged to promote activation/installation. Such pings were used to promote ENX but not custom EN apps. When this situation was pointed out, Google immediately offered the same adoption promotion pings for custom apps as for ENX, while Apple declined to do so.

A number of players, including but not limited to Apple, argued that the key to persuading people to adopt lay in ensuring the privacy of the system. Fortunately, decentralized protocol designs offer powerful solutions to the privacy problem. Centralized approaches send information about who went where with whom to some central database. This breach of privacy, while clearly facilitating contact tracing, also poses risks of abuse as part of a mass surveillance state. Under a decentralized protocol, information about a user is stored on that user's device, and to some degree the devices of those they were in contact with. Users can be given autonomy over the use of data on their own devices, so that e.g. it cannot be accessed without consent, and can be deleted at any time. Far from threatening privacy, apps using decentralized protocols are among the safest apps on users' smartphones.

Interestingly, privacy-invasive schemes, such as QR code check-in presence tracing in Australia, were well accepted and widely used (self reports of 61.9% always checking in either digitally or on paper, 26.3% mostly, 4.7% sometimes, 3.3% occasionally, 3.8% never [17]). This achieved acceptably low failure rates far superior to any EN implementation, even with enforcement left to the venue and/or peer pressure. Indeed, supported by celebrity-driven PR pushes, 40% had downloaded or were willing to download even the ineffective but convenient Luca QR code check-in app in Germany [18, 19]. Perhaps these much higher adoption rates for presence tracing than for proximity notification are because what was recorded was considered public information that a user was in a public space, rather than also capturing who had interacted with whom even in private. A simpler explanation is that the QR code check-in system is easy to understand as equivalent to writing down your name and number; decentralized privacy-preserving schemes, by being harder for the public to understand, run the paradoxical risk of decreasing rather than increasing trust and hence adoption.

Trust in government predicts adoption more strongly and consistently than privacy [15, 20-23]. While 20-25% respondents to cross-sectional surveys do cite privacy concerns as a reason for not installing a contact tracing app [23, 24], a longitudinal study found no causal relationship (although it did for health concerns and social norming) [25], and a focus group found that health concerns were more important than privacy considerations [26].

High trust in government in general [27], as well as in the government's response to the COVID-19 pandemic [28], was critical to Singapore's high adoption. When trust in governments is low, it might help if more trusted entities, such as primary care providers [29], were used to promote adoption. Singapore's explicit policy to relax certain pandemic restrictions once adoption rose from 50% to 70% might have also helped drive adoption, together with the scheme's high profile as a clear government priority.

Privacy is not the only aspect of trust salient to potential users; many also wanted to know that an app was effective before they downloaded it. Even if they decided to take a chance at first and install it, they wanted subsequent reassurance that it had proved effective, in order to keep it active on their phones. Multi-functional apps, like the NHS COVID-19 app and Corona Warn app, can also provide reassurance via interaction that there is a point to maintaining the app on the user's phone. While it is clearly important to hire good marketing/public relations professionals to promote adoption, it is just as important that scientists and engineers do not make product decisions in isolation from their impact on marketing, because achieving high adoption is part of the science of making them effective.

Distinct from the scientific study of app effectiveness in stemming transmission [30], is the study of viral transmission itself, to learn its incubation period, infectious period, and mode of transmission [31]. An obstacle to such study was that some "privacy-first" rhetoric rejected making the study of SARS-CoV-2 transmission an aim of the scheme, even if the science could be done in a manner that preserved privacy.

The only reliable way to learn about transmission to and from humans is to observe it, either in human challenge trials (directly observing incubation periods, and also the infectious period if the study includes transmission to co-housed animals), or in natural circumstances (i.e. via contact tracing). Given ethical concerns about the former, an unacknowledged corollary of rejecting the use of digital schemes to study transmission is that studies must be performed via manual contact tracing (i.e. in a more privacy-invasive way), or not at all. Unfortunately, even the more basic functions of manual contact tracing were quickly overwhelmed, leaving little capacity for the more intensive investigations required to study transmission dynamics. Not using digital approaches to fill this void was a lost opportunity. It left early manual contact tracing studies as the only source for basic transmission parameters, even after the incubation time and infectious period of SARS-CoV-2 were suspected to have shifted due to viral evolution and/or immunity.

Even an app whose failure rates across the six points were too high to substantially reduce transmission could have been enough to generate valuable information about evolving incubation periods and infectious periods in close to real time [31]. Apps can also help with epidemiological surveillance; e.g. the Indian government combined location (with a random error added for privacy; EN apps were banned from accessing location) and symptom information to identify emerging hotspots prior to seeing spikes in test positivity. Singapore used the numbers of detected contacts per case to inform changes in social distancing policies. Support for the scheme, both from the public and from public health, might have been shored up by evidence that it was at least doing good for science and/or disease surveillance, and this might have created a virtuous circle of adoption.

A prevailing narrative described the technological choices in terms of a trade-off between privacy and effectiveness, in which any collection of more data can increase perceived privacy risk and hence reduce public acceptance [30]. We believe this framing is harmful, especially when it comes to promoting app adoption; e.g. the public might then conclude that because the tech is private, it cannot be effective. As for substantive risks beyond those of perception, with flexibility and creativity, effectiveness can be achieved by a decentralized, privacy-preserving design that does not risk expanding mass surveillance. With sufficient ingenuity, there may be no need for compromise at all, which makes the privacy vs. effectiveness tradeoff narrative misleading. "Privacy first" should therefore not be used as grounds to support the status quo to the point of refusing to engage in ongoing dialog regarding how to safely proceed with iterative improvements to effectiveness, its measurement, and the gathering of broadly valuable scientific and/or epidemiological surveillance data. Given that harms from the technology are not limited to privacy concerns [32], a privacy-preserving design does not relieve governments of their duty to monitor effectiveness [33].

Furthermore, privacy is merely a component and a means to the more important end goals of autonomy and protection from abuse. The latter need not be a serious issue, because a well-designed, decentralized solution can offer both effectiveness and protection from abuse. However, trade-offs between effectiveness and autonomy are real, i.e., coercion can increase effectiveness while autonomy can weaken it. The option to maintain privacy while extracting benefit from the system is an important but not the only aspect of autonomy.

While the privacy approach of EN provides excellent protection from abuse, it actually limits autonomy by denying users the right to share their data with public health if they wish to. This became clear when one of us [JP] proposed a small change to the EN protocol that could convert it from proximity notification to proximity tracing, without significantly raising the danger of misuse. Described further in Section 3, this modified protocol would allow users who had both received a notification and tested positive to share more data with public health, given explicit consent. By helping manual contact tracers to do their job of pairing cases and identifying superspreading events, this change would presumably have generated more buy-in from public health authorities (discussed in Section 5). More buy-in from the local public health authorities responsible for manual contact tracing could have set up a virtuous circle by which they then in turn promote adoption in the community. This small modification was rejected out of hand by Apple as being a threat to privacy, despite the fact that any privacy loss is triggered by users exercising autonomy over the use of their data. Contributing to this dynamic is the fact that privacy is part of Apple's brand, aligned with broader deployment of anti-tracking as a business strategy [34-36]. A preference for privacy over autonomy, by disallowing the opt-in sharing of more information with public health, has also been expressed by some academia-based protocol designers [37].

The slogan "privacy first" is an odd one. We believe that autonomy and protection from abuse are better-framed goals than privacy. Furthermore, the best way to put privacy first is self-evidently to have no digital system at all, raising the possibility that "privacy first" schemes were designed primarily to displace more invasive options, with stemming transmission almost an afterthought. For each proposed scheme, one should assess the nature and scope of any vulnerabilities vs. the anticipated impact on disease transmission.

## 3. Detecting and evaluating exposure

Point of failure #3 concerns the need to record an interaction at which viral transmission took place, but viewing it this way is not sufficient. Total success could trivially be achieved by a general stay-at-home order, which is equivalent to telling everyone that they are potentially exposed. This illustrates how it is also important to avoid false positives, i.e. alerting people who are not infected. The primary purpose of proximity/presence notification is best seen as identifying and alerting people at sufficiently high risk of infecting other people, such that they should change their behavior in response to that information. This benefit applies only to secondary cases who would not otherwise know of their exposure in a timely manner, i.e. it excludes household contacts and other social networks that perform rapid do-it-yourself tracing. A major advantage of digital schemes is their complementary ability to notify strangers.

One aspect of this is to detect any contact at all. With core (i.e. regular) Bluetooth, iPhone to iPhone communication does not work in the background. The need for Apple's co-operation setting up EN was to get around this serious obstacle. Apple could have chosen to enable iPhone to iPhone communication in the background the same way that they already allowed background communication between iPhones and Androids - this would have allowed app developers to implement their own protocols. Indeed, India launched the Aarogya Setu app using core Bluetooth rather than EN, given that iPhones are rare in India compared to Androids. Singapore stuck to its own BlueTrace protocol (based on core Bluetooth), given the sacrifice of epidemiological utility that switching to EN would have implied [38], eventually leveraging Android-iPhone background communication into a gossip protocol that effectively allowed iPhones to see each other in the background whenever an app-using Android was also present [39]. Other non-EN jurisdictions such as France pivoted away from proximity notification and toward presence tracing. In Australia, State-specific centralized presence tracing apps were the dominant response, while a federal app based on TraceTogether, despite a respectable early download rate, languished once it became apparent that it failed at other steps.

Lack of interoperability among jurisdictions can also interfere with the detection of contact. E.g. if each person is far from home for 5% of their interactions (e.g. being away for 18 days per year), the corresponding contact detection failure rate will be ~10%. When thus viewed quantitatively, the considerable attention devoted to this issue seems disproportionate relative to other sources of failure. As an alternative to immediate global standardization, metropolitan or other highly interconnected areas spanning multiple jurisdictions (e.g. D.C. combined with adjacent counties in other U.S. States, or the Navajo Nation spanning multiple U.S. States) could choose to roll out a joint product. Multiple apps need not interfere with one another, making this an option for some commuters. Common servers were set up to enable communication between the EN implementations of different U.S. States, and another between those of different European nations [40], but despite the attention to this issue, global interoperability was never achieved. Perhaps the best model for a new pandemic is to allow divergence in the early stages as part of a process of innovation at the cost of interoperability, to learn from the different experiments performed by different jurisdictions, and then to use the desire for interoperability as the catalyst for switching from lower performing systems to higher performing through app updates. An advantage of postponing interoperability is to ensure that premature standardization does not suppress innovation.

Beyond detecting exposure at all, doing a better job in assessing the infection risk posed by a given exposure can be seen as an ethical imperative, given the harms caused by unnecessary quarantine [32]. Better risk analysis means more precise resolution as to which individuals pose how much statistical risk to others on a given day, and using a risk threshold to trigger notification. In other words, good risk analysis means improving the ROC curve with respect to infection, and choosing a socially optimal point along the ROC curve.

Under reasonable assumptions, the socially optimal approach is to alert those who are significantly more likely to infect others than is the average member of the population not already in quarantine or isolation [3]; for the NIH COVID-19 app, notified individuals were 2- to 20-fold more likely to subsequently report a positive test [41]. Under the assumption that some form of regulation keeps the geometric mean of R(t)~1 while waiting for a vaccine, the total value of both manual contact tracing and digital notification schemes is highest at low case prevalence [3], when fewer individuals need to quarantine to achieve the same population benefit. Digital contact tracing also becomes more important when social restrictions are few, and interactions among strangers thus more common. Unfortunately, in many jurisdictions, attention to all pandemic measures tended to rise and fall with case counts, rather than fluctuating between a focus on contact tracing during lulls and on population-wide measures when case counts were high. This interfered with effective, planned implementation.

One obstacle to good risk analysis was public health guidance that ignored evidence for airborne transmission in favor of the droplet theory that physical distancing of 2 meters (or 1.5 meters, or 6 feet) was effective protection. As a result, significant tech research focused on calibrating EN Bluetooth settings to a threshold of 2 meters or similar [42-45]. A number of unpublished analyses suggest close to superimposable ROC curves regardless of Bluetooth settings. Superior distance assessment via Bluetooth could likely have been achieved by using all 3 Bluetooth channels instead of one, and/or using the median attenuation instead of the mean (decisions made within the EN API and hence not available to app developers). Ultra-wideband or ultrasound would assess distance still more precisely. But none of these tweaks change the fact that distance only moderately predicts infection risk, and proportionate effort was not invested in integration with predictors such as local $CO_2$ levels. Nor was information such as the time or location of exposure provided to users, enabling them to integrate information about masking and ventilation into their personal assessment of the risk, despite that information being present on the user's phone. Not allowing this, even with the permission of the primary case, is another example of privacy being put before autonomy.

After coming up with "good enough" Bluetooth settings, a better approach would have been to deploy an app that collected data both on the parameters of exposure, and on whether it was followed by infection. This data could have informed risk settings better than any experiment on Bluetooth-distance relationships. We note that a protocol might flag an individual as high-risk on the basis of an exposure other than the one that infected them; real-world calibration will include such cases, potentially increasing impact beyond that of causal connections.

An ongoing process of data collection and risk calibration would have taught us much about transmission, e.g. any changes to the timing of infectiousness as new strains appeared, and as individuals acquired prior immunity. This is important because exposure dose, and hence infection risk, depends not just on the physical characteristics and duration of an encounter, but also on how infectious the primary case is at that moment. The main data used to assess this was the date of symptom onset, or if asymptomatic, the date of first positive test. The infectiousness window was initially estimated to run from around 2 days before to 5 or more days after symptom onset [46]. Unfortunately, a bug in the code of the original analysis concealed the earlier onset of infectiousness; while this was rapidly identified and corrected [47, 48], public health guidelines, describing for which days the primary case should be considered infectious for contact tracing purposes, did not change accordingly. Nor did they change with the emergence of new SARS-CoV-2 variants, nor for reinfection or breakthrough cases, for which the same kind of intensive manual contact tracing studies were not conducted to estimate this window. Other information about infectiousness could in principle come from Ct count (which was not however reported in standard laboratory SARS-CoV-2 PCR testing protocols) or connection to a superspreading event, but neither was used in practice.

EN version 1 allowed the app of the primary case to assign one of 8 levels of infectiousness to each day during which others might have been exposed. This integer is the only metadata regarding the primary case that is available to the exposed contact. EN v2 enabled some improvements to risk analysis by lifting the previous 30 minute cap on exposure reported by the contact's phone, but reduced the number of levels of infectiousness down to 2 (although Germany repurposed other bits of information to reclaim the use of 6 [49]). When asked to make the infectiousness metadata easier to use for both risk analysis and scientific study, specifically to propagate the timing of exposure relative to symptom onset / test date from the server to the exposed contact, Apple replied that they couldn't do it because of a vulnerability whose details had been worked out by Google but which they could not remember.

Beyond EN, some jurisdictions acknowledged airborne transmission to the point of performing not just proximity tracing but also presence tracing. In corresponding presence tracing/notification schemes, people going to venues such as restaurants would either write down their name and contact details, or else use an app to check in by scanning a QR code, or for Singapore's SafeEntry they could also use a Bluetooth-based sensor (to detect a TraceTogether app or token), or scan a barcode. With a low risk threshold, e.g. a jurisdiction with a zero COVID policy, all individuals who were present at the same venue as an infectious individual can be alerted.

A better strategy for a higher risk threshold is to focus on "cluster busting", i.e. prioritizing those venues at which at least one transmission event is already known to have occurred. These generally correspond to gatherings at which the primary case was highly infectious, ventilation was poor, and/or there were other risk factors such as speaking/singing/exercising and lack of mask use. There are compelling arguments for the effectiveness of such "backward" tracing for pathogens such as SARS-CoV-2 [50-52]. This requires cases to be linked.

A decentralized protocol like EN could have contributed to cluster busting, had the app of a user who was first exposed then tested positive been allowed to upload the cryptographic key responsible for triggering the exposure. EN chose to keep this key sequestered within the operating system, with no option to notify the server. If a key associated with a transmission event were uploaded, the infectiousness score of that key could then be significantly upgraded on the server, leading to notifications given briefer exposures to the possible superspreading event, and also more strongly worded notifications. If users were additionally given the option of sharing their identity with public health, paired with the cryptographic key that served to link them, then both members of the pair could be prioritized for interview and the location of the superspreading event identified to assist manual contact tracers.

## 4. Test access

No form of contact tracing, digital or manual, will work unless we are able to identify who is likely infected, and thus whose contacts should be notified/traced. While some countries such as South Korea did better, in many countries, SARS-CoV-2 tests were slow to roll out, testing capacity was quickly overwhelmed, test shortages persisted for a surprisingly long time, and not all symptomatic individuals sought testing [53-55]. It is obviously critical to improve pandemic preparedness in this regard, independently of its relationship with digital notification/tracing.

Improving allocation of scarce testing resources, by identifying which individuals were most likely to test positive, was a primary use case for India's Aarogya Setu, with the highest risk individuals having positive predictive values above 40%. To achieve this, Aarogya Setu used a sophisticated risk algorithm to trace not just direct contacts, but contacts of contacts and so on (multiple degree tracing) in a risk-consistent manner, faster than testing could keep up [56]. In EN, functionality for contacts of contacts was limited, relied on self-reporting, and was never used. Another important factor for allocating scarce testing resources is utility for onward contact tracing/notification, e.g. by abandoning samples more than a certain number of days past symptom onset in order to rapidly turn around tests whose results will be more actionable.

For some pandemics, there may be one or more "hallmark" symptoms (such as loss of taste and smell) that are sufficiently distinctive to warrant presumptive diagnosis. In other cases, symptoms might be suggestive but less definitive. Strategies for triggering notifications given contact with an unverified case, in the absence of test results, overlaps with the next point of failure, namely how to manage verification of positive test results.

The CoEpi app [57] attempted to launch on the basis of symptoms alone, using core Bluetooth (it was designed prior to EN). It was rejected from the Apple Store on the basis of lack of public health involvement, with the suggestion to seek public health support from a "city or county". Its resubmission with explicit public health support and endorsement from a county in Washington State (from a different non-profit account, because Apple suspended the primary developer's account without explanation), was also rejected. Apple representatives told developers that unwritten and non-public rules stood in the way, and that it was the desire of Apple to not have "competing" approaches.

When test turnarounds take several days, but each day is critical, it might be helpful to issue "preliminary" notifications on the basis of exposure to an exposed, symptomatic individual awaiting test results, to be converted later to a confirmed exposure. The same principle can be expanded deeper into a social network. We note that recursive protocols can incidentally achieve some of the functionality of backward contact tracing, showing up as multiple social network paths via all attendees to the index case.

## 5. Triggering notifications

Following a positive diagnosis for SARS-CoV-2 in a primary case, speed is of the essence in order to notify secondary cases before they transmit onward to others. The simplest option is to allow any app user to self-attest that they are infected. An issue with this is the potential for abuse. E.g., prior to an election, a coordinated set of individuals could intentionally socialize widely in a setting that tends to vote a particular way, e.g. a college campus, and then falsely report positive diagnoses timed to trigger quarantine on the day of the election. This was a specific concern for the U.S. November 2020 elections. Disruption of essential workers at core infrastructure might also be achieved through targeted attack.

In this light, and to ensure trust in the system, all EN jurisdictions launched with a requirement that central authorities set up a system to ensure that only verified positive test results could be used to trigger notifications. Many jurisdictions initially relied on manual contact tracers to issue a time-sensitive secure code over the phone to individuals who tested positive. This led to extremely high rates of failure, given that case investigation was overwhelmed. EN imposed extra work on case investigators in issuing verification codes, but provided nothing back to them in return e.g. by helping link cases into transmission chains, as could be done by the protocol tweak described above. Even when the code was successfully delivered and entered into an app, this tended to happen with a significant delay. This defeated much of the purpose of a digital scheme, whose motivation was to make contact tracing faster [58]. Primary care can similarly become overwhelmed, and in one controlled, non-overwhelmed setting, voluntary adherence even once a code is provided was only 64% [14]. In contrast, Singapore's solution was fully integrated with both testing and manual contact tracing, speeding up the latter from 4 days to less than 2 days [59], and switching to primarily automated alerts only after case counts rose during the Delta wave in late 2021.

Google and Apple insisted on having only one EN app per country, or in the case of the U.S., one per State. However, test and trace programs are run by counties and tribal nations in the U.S., by provinces in Canada, by cantons in Switzerland, etc. This disconnect did not encourage buy-in by the local public health authorities charged with distributing secure codes; app design choices were made by a different level of government than the level responsible for implementing test and trace policies.

Presumably Google and Apple preferred to limit the number of relationships they needed to maintain, and of EN apps whose code they needed to review. However, this could also have been solved via a flexible global app (or an ecosystem of several at least partially interoperable apps); Google and Apple would only need to deal with one or several app developers, who in turn would deal with the customizations requested by the various jurisdictions. Such an ecosystem began to spontaneously appear through players such as NearForm, PathCheck, and WeHealth. A bottleneck in many U.S. states was the slow process of government procurement to pay such players. One reason many U.S. states opted for ENX was that because no payment was required, it bypassed delays in the procurement process, despite the fact that running ENX and associated verification code distribution still generally required a State public health authority to internally dedicate full-time staff. With the advent of ENX, Google and Apple ended up needing to maintain more relationships than they would have with private sector middlemen.

Given the difficulties in giving verification codes out by phone, most U.S. implementations, beginning with the State of Colorado, shifted to an SMS-based system. Positive test results reported to the State were collated, and cases were then, in batches, sent an SMS with a deeplink that acted as a verification code. Vague language and a few decoy SMSs were used to accomplish HIPAA compliance. There was still some post-testing delay associated with reporting and collating, some SMSs were caught by spam filters, some were lost to the SMS-delivery network during peak usage periods, and the system was confusing for recipients who had never heard of EN. But it was both faster and had a higher success rate than having a manual contact tracer issue a code over the phone. The ratio of #claims/#cases (measuring a combination of failure points 1 and 5) went up, e.g. from 1.8% to 9.6% in the State of Washington (with the caveat that both are upper bounds on success because the total neglects delays in receiving codes, and includes codes claimed by individuals who installed EN only after receiving the deeplink) [60].

An alternative way to issue verification codes was to integrate with testing, i.e. with the healthcare system rather than the public health system. The best solution is to make the app into a test result delivery system, i.e. to link test samples to QR codes, with the app knowing the IDs of its user's test and checking a server for matches. This shortens the time from a sample testing positive in the laboratory to notifications being triggered, but still requires the user to check their phone for a positive test result, and consent to notifications. In Singapore, the median time from SMS notification of test results to consent to upload data was <30 minutes, with a consent rate of 70-80%. EN was eventually modified to allow a still better pre-authorization workflow, whereby consent could be given at time of testing, and notifications were sent as soon as the app became aware of the positive test result, with no further user input required. This system was used in Germany. Germany also offered free supervised antigen tests: the unvaccinated required a recent verified negative result to permit access to some venues, while positive results could be verified by the same system for use in EN.

Integration with testing was easier for the NHS COVID-19 app than for many others, given an already-centralized system. However, it was also possible elsewhere; Germany integrated its app with over ten thousand different test providers. The speed and convenience of getting a PCR test result back through the app helped prompt download, although some users opted only to use this feature and not to trigger notifications. It is important to investigate the cause of this refusal, e.g. did it stem from distrust regarding anonymity, were people trying to avoid pushing friends and family into uncompensated quarantine, or was it simply an instinctive "no" under time pressure in a stressful situation, for which a different workflow might have elicited a different decision?

Two EN pilots at U.S. universities explored integration with campus testing programs, which provided rapid test turnaround prior to adequate testing being available to the general public. The University of Alabama linked test results to phone numbers, then allowed individuals to use their phone number as their verification code through use of a one-way hash. The University of Arizona gave out verification codes as part of the online portal from which test results were distributed, with 25% of cases claiming a verification code [13]. As part of an (abortive) expansion to the State at large, Application Programming Interface (API) integrations were created not just with the State's largest test providers (LabCorp, SonoraQuest and others), but also with Doximity, an app believed to be used by 70% of the State's doctors. Surveys show higher trust in healthcare than in public health [29, 61], making doctors' involvement potentially useful.

In retrospect, all these schemes were too complex to work reliably from the outset, at least in most jurisdictions. Better might have been to allow self-attestation of positive test results, or even self-attestation on the basis of symptoms alone at times during which tests are in short supply (see section 5 above). To avoid the potential for malicious use, unverified reports could trigger notifications with different messaging: not requiring quarantine, merely warning the recipient, and including the fact that the person who exposed them had self-attested to their infected or symptomatic status. Once post-exposure quarantine was relaxed and home testing became the norm, many EN jurisdictions switched to self-attestation.

All the same difficulties arise with presence tracing. Many QR code check-in schemes required manual contact tracers to identify locations of interest in order to trigger app notifications. Like other forms of manual contact tracing, including pen-and-paper check in at venues, this process was often overwhelmed. While fully automated systems would have been faster and more reliable, they were not permitted for apps that also ran EN. New Zealand implemented this functionality early on, i.e. users who tested positive were able to upload the set of QR codes they had scanned to public health. However, when the NHS COVID-19 app attempted to follow suit, its app update was rejected [62], and New Zealand was also forced to make changes. Google and Apple allowed EN app users to store a local copy of their check-in history on their phone and to read it over the phone to manual contact tracers, but they did not allow an opt-in upload button; again, a rejection of user autonomy on the grounds of privacy. They did permit an alternative automated QR code check-in system in the German EN app that was designed to never be linked to identities, but because this system did not satisfy laws in most German States requiring the collection of identity information, this limited its adoption.

Finally, many systems assumed that once an individual tested positive, they would enter isolation such that only past contacts and not future contacts would need to be notified. This turned out to be over-optimistic. We recommend that future systems anticipate non-adherence, with daily prompts of "Did you succeed in isolating today?" leading to the option upload new data in order to anonymously notify that day's contacts, in addition to contacts from the initial upload.

## 6. Behavior change

To stem transmission, tracing/notification needs to result in behavior change by secondary cases. Rates of quarantine can be low, estimated as 28% for the asymptomatic in Norway adhering for at least one day [63], and 11% in the U.K. for quarantine adherence on all recommended days [64] (albeit much higher in the app-using subset [65]). Quarantine adherence 40% among the app-using subpopulation in the Netherlands [61]. Adherence increases with trust in governmental response to a pandemic [66].

Short of outright coercion, quarantine adherence might be improved by paying people, whether paid directly by the government or via employer mandates. Adherence might also be higher if, out of respect for their time, rigorous risk analysis (including the use of negative tests) were used to reduce quarantine duration to a minimum, with a risk threshold set on a rational basis. Some European countries gave quarantine pay for individuals traced by contact tracer but not those notified by EN. It is important to make such choices rationally, i.e. to estimate the positive predictive value of an EN with given characteristics, and to treat individuals with the same risk of infecting others similarly, regardless of the mode of risk detection.

There can still be substantial utility to issuing low exposure warnings to individuals whose positive predictive value does not warrant quarantine. Short of full quarantine, more modest behavior changes also help stem transmission, e.g. testing and isolating if positive [67], being more alert for symptoms and isolating if they appear, mask-wearing, or avoidance of large groups and/or clinically vulnerable individuals. However, adherence to even these more modest requests is also far from universal, e.g. in the State of Washington, only 40% of the subset of EN recipients that responded to a survey intended to get tested and 67% intended to watch for symptoms; rates were 58% and 84% on the smaller subset that also replied to a follow up survey about actual behaviors [68]. With Paxlovid or other antivirals most effective when taken early, we note that even if notification fails to stem transmission, it can be of direct benefit to an individual who makes no changes other than monitoring for symptoms, then testing and treating if positive. Adoption rates (points of failure 1-2) might be higher if this direct benefit were stressed, rather than relying on altruistic motives.

Even without complexities such as recommending quarantine for some but not all exposures, effective communication of "next steps" at a low reading grade level is surely important for adherence. In contrast, some EN implementations launched with quarantine recommendations that did not specify an end date (leaving users to assume it began with the date of notification receipt rather than the unknown date of exposure), or gave the date of exposure and left the user to perform the calculation of quarantine end date.

While the purpose of this piece is to present lessons learned, the better to inform responses to the next airborne pandemic, we note that the SARS-CoV-2 pandemic remains unpredictable in its evolution, and that EN has not yet been sunsetted in all jurisdictions. In that light, we note that the end of quarantine need not mean the end of EN. If notification caused individuals to mask, avoid the most vulnerable, and/or test then isolate if positive, then the tech would still be doing something useful, should low enough rates of failure at stages 1-5 be achieved.

## Governance for effectiveness

To contain a pandemic, i.e. to achieve R(t)<1 through digital tracing/notification, an implementation must keep all six failure rates low, e.g. below 20%. This will require good governance, for which it can be instructive to learn from the best implementations so far. The NHS and Germany had among the best EN implementations, even if their failure rates remained far higher than this benchmark. Singapore's non-EN implementation was more successful, with <5% failure rate for 1-2, and acceptably low failure rates for 4 and 6. We hope that more scientific data might eventually emerge from it.

It is notable that both the German and NHS COVID-19 app projects engaged not just software developers, privacy experts, and applied public health practitioners, but also well-respected academic epidemiologists with significant track records of directly related research, and that these epidemiologists had substantial (but not sole) influence over decision-making. They might be in the best position to balance ambition of scope (to achieve R(t)<1 not "we did something") with realistic expectations about how things will play out on the ground. This follows a more general pattern during the COVID-19 pandemic in which applied public health institutions such as the CDC performed badly, but many universities and research institutions such as the Robert Koch Institut performed relatively well, as did biosecurity-run initiatives such as Operation Warp Speed, and the pharmaceutical industry when given a good incentive structure.

The fact that iPhone to iPhone low energy Bluetooth communication did not initially run in the background meant that some response from Apple was needed for effective Bluetooth-based proximity tracing/notification using iPhones. Once that response took the form of EN, countries that did not capitulate to Apple's extensive conditions for access often ended up focusing on presence tracing/notification systems instead (e.g. France, Australia), unless they used a gossip protocol and/or issued alternative hardware (Singapore), or had few iPhones in their country (India). Apple's ability to exert power over EN protocol design and over API access substantially restricted the scope for innovation among individual apps. Whatever the best form of governance is, few would argue that it is best done behind the closed doors of a tech company. Given that Apple exerted its power to dictate terms and limit innovation, it is a tragedy that Apple did not also use its power to enact an opt-out system. For example, the NHS COVID-19 app is estimated to have saved 10,000 lives in its first year [41], and it is estimated that for every 1% increase in uptake, it could have reduced cases by a further 1-2% [2].

Future success will require measurement of failure rates at all six points, and rapid on-the-fly adaptations to improve them. Ideally, we would invest now in these technologies, perhaps within island nations or other close-knit communities as test cases, to iteratively improve systems as part of pandemic preparedness, while at the same time attempting to reduce SARS-CoV-2 and potentially also influenza transmission in the short term. An iteratively improved technology is more likely to be successfully deployed should a new pathogen begin transmitting between humans, one that combines SARS-1 or H5N1 mortality with SARS-2 presymptomatic transmission. In the absence of current investment, we still hope this document helps kickstart the design of effective strategies at such a time.

### Acknowledgements
We thank Ryan Barrett, Justus Benzler, Dana Lewis, Ron Rivest, Curran Schiefelbein, and Tina White for helpful comments on the manuscript.

### Conflicts of Interest

Joanna Masel and James Petrie have an equity interest in WeHealth Solutions PBC, which distributed an EN app to Arizona and Bermuda. Andreas Gebhard is CEO of Forward Momentum, LLC, an industry-agnostic consulting firm which may at times serve clients in the software, hardware or public health space.

### Abbreviations
API: Application programming interface
EN: Exposure Notification

ENX: Exposure Notification Express
NHS: National Health Service